\documentclass[]{interact}

\usepackage{epstopdf}
\usepackage[caption=false]{subfig}

\usepackage[numbers,sort&compress,merge]{natbib}
\renewcommand\bibfont{\fontsize{10}{12}\selectfont}

\usepackage{graphicx}
\usepackage[tight-spacing=true]{siunitx}
\usepackage{multirow}
\usepackage{amsmath,amssymb,verbatim}
\usepackage{bm}

\theoremstyle{plain}

\theoremstyle{definition}

\theoremstyle{remark}

\newcommand{\hdplus}{{$\text{HD}^+ $}}
\newcommand{\mpme}{$m_{\rm p} / m_{\rm e}$}
\newcommand{\mdmp}{$m_{\rm d} / m_{\rm p}$}
\newcommand{\mdme}{$m_{\rm d} / m_{\rm e}$}
\newcommand{\Ar}[1]{$A_{\rm r}({\rm #1})$}
\newcommand{\ArM}[1]{A_{\rm r}({\rm #1})}
\newcommand{\ArMCODATA}[1]{A^{\rm CODATA2018}_{\rm r}({\rm #1})}
\newcommand{\pHe}{$\bar{\text{p}}\text{He}$}

\newcommand{\vTP}{$(v,L)$:~$(0,3) \rightarrow (9,3)$}
\newcommand{\vRot}{$(v,L)$:~$(0,0) \rightarrow (0,1)$}
\newcommand{\vVib}{$(v,L)$:~$(0,0) \rightarrow (1,1)$}

\begin{document}


\title{Extraction of spin-averaged rovibrational transition frequencies in \hdplus{} 
	for the determination of fundamental constants}

\author{
	\name{Jean-Philippe Karr\textsuperscript{a,b}\thanks{CONTACT J.-Ph. Karr. Email: karr@lkb.upmc.fr} and Jeroen C.~J. Koelemeij\textsuperscript{c}\thanks{CONTACT J.~C.~J. Koelemeij. Email: j.c.j.koelemeij@vu.nl}}
	\affil{\textsuperscript{a}Laboratoire Kastler Brossel, Sorbonne Universit\'e, CNRS, ENS-Universit\'e PSL, Coll\`ege de France, 4 place Jussieu, F-75005 Paris, France; \textsuperscript{b}Universit\'e d'Evry-Val d'Essonne, Universit\'e Paris-Saclay, Boulevard Fran\c{c}ois Mitterrand, F-91000 Evry, France;
		\textsuperscript{c}LaserLaB, Department of Physics and Astronomy, Vrije Universiteit Amsterdam, De Boelelaan 1081, 1081 HV Amsterdam, The Netherlands.}
}

\maketitle

\begin{abstract}
We present a comprehensive analysis of all currently available high-accuracy frequency measurements of rotational and rovibrational transitions in the hydrogen molecular ion \hdplus{}. Our analysis utilises the theoretically calculated hyperfine structure to extract the values of three spin-averaged transition frequencies through a global linear least-squares adjustment that takes into account theory-induced correlations between the different transitions. We subsequently use the three spin-averaged transition frequencies as input data in a second adjustment which employs precise theoretical expressions for the transition frequencies, written as a function of the proton, deuteron and electron relative atomic masses, the Rydberg constant, and the proton and deuteron charge radii. Our analysis shows that the \hdplus{} data may significantly improve the value of the electron relative atomic mass and the proton-electron mass ratio, in particular if combined with recent high-precision measurements of particle atomic masses and mass ratios obtained from Penning traps.
\end{abstract}

\begin{keywords}
Deuterated hydrogen molecular ion; precision rotational-vibrational spectroscopy; ab initio molecular theory; molecular hyperfine structure; fundamental physical constants
\end{keywords}

\section{Introduction}
Fundamental physical constants are determined from precision measurements, often performed on single particles or simple atomic quantum systems, by adjusting the values of the constants such that theoretical predictions match the experimental observations as closely as possible. Such adjustments, implemented via least-squares optimization of the constants of interest with respect to large bodies of input data, are carried out by standards organizations such as the Committee on Data for Science and Technology (CODATA) and the Atomic Mass Data Center (AMDC). 

For example, the most precise determinations of the Rydberg constant, $R_\infty$, and the electric charge radii of the proton, $r_{\rm p}$, and the deuteron, $r_{\rm d}$, have been obtained from laser spectroscopy of both electronic and muonic hydrogen and deuterium, complemented by high-precision quantum electrodynamics (QED) theoretical predictions (see Ref.~\cite{Tiesinga2021} and references therein). By contrast, the relative atomic masses of particles like the electron and light nuclei such as the proton and deuteron have traditionally been determined using Penning-trap experiments and (for the electron) theoretical calculations of bound-state $g$-factors; for recent examples see Refs.~\cite{Sturm2014,Heisse2017,Heisse2019,Rau2020,Fink2020,Fink2021}. In 1976 Wing et al. already recognised that vibrational spectroscopy of hydrogen molecular ions, such as H$_2^+$ and \hdplus{}, could also be used for determinations of particle mass ratios provided that both the measurements and the theory of these three-body, two-center quantum systems would be sufficiently accurate~\cite{Wing1976}. This concept was extended to laser spectroscopy of antiprotonic helium, \pHe{}, by Hori et al. who combined their results with precise ab initio theory for a meaningful determination of the antiproton-electron mass ratio which under the assumption of charge, parity and time reversal (CPT) invariance must be the same as the proton-electron mass ratio, \mpme{}~\cite{Hori2006,Hori2011,Hori2016}. As a result, \pHe{} data were included in the 2006 and 2010 CODATA adjustments where they contributed primarily to the value of the electron relative atomic mass, \Ar{e} ~\cite{Mohr2008,Mohr2012}. However, the 2014 and 2018 CODATA adjustments no longer considered \pHe{} after \Ar{e} was determined using Penning-trap measurements and QED theory with over an order of magnitude smaller uncertainty~\cite{Sturm2014}. 

Over the past two decades, advances in laser and terahertz spectroscopy of \hdplus{} stored in radiofrequency (rf) ion traps and sympathetically cooled by laser-cooled ions, considerably improved the accuracy, culminating in the measurement of various rovibrational transition frequencies with uncertainties in the low parts-per-trillion (ppt) range~\cite{Alighanbari2020,Patra2020,Kortunov2021}. Meanwhile, the precision of theoretical QED predictions of relevant transition frequencies in \hdplus{} improved to below 10~ppt, such that the uncertainty of theoretically predicted frequencies is now limited by (primarily) the uncertainty of the 2018 CODATA value of \mpme{}~\cite{Korobov2017,Aznabayev2019,Korobov2021}. As anticipated by Wing et al.~\cite{Wing1976} and more recently also by Karr et al.~\cite{Karr2016}, the availability of theoretical and experimental transition frequencies with uncertainties in the low-ppt range enabled a determination of \mpme{} with a precision of about 20~ppt~\cite{Alighanbari2020,Patra2020,Kortunov2021}. These values represented not only the most precise determinations of \mpme{} to date, but they were also found to be consistent with a recent precise measurement of the proton relative atomic mass, \Ar{p}~\cite{Heisse2017,Heisse2019}, that deviated significantly from determinations of \Ar{p} included in the 2018 CODATA adjustment~\cite{Tiesinga2021}. Furthermore, several researchers have pointed out that results from \hdplus{} and \pHe{} form a link between 'atomic' fundamental constants such as $R_\infty$, $r_{\rm p}$ and $r_{\rm d}$, and particle mass ratios~\cite{Karshenboim2017,Antognini2022,Antognini2022arXiv}.

The recent advances in \hdplus{} experiments and theory and their demonstrated potential for the determination of fundamental constants warrant a closer investigation to prepare available \hdplus{} data for inclusion in future CODATA adjustments. One crucial issue is that, in general, atomic and molecular spectroscopic data include line shifts due to hyperfine interactions, whereas CODATA adjustments have historically considered only energy differences between hyperfine centroids. This implies that experimental \hdplus{ }spectroscopic data must undergo a preparatory analysis in order to remove hyperfine shifts from measured rovibrational transition frequencies. Here, we will use the term  'hyperfine components' to indicate rovibrational transitions between individual hyperfine levels. In principle, the spin-averaged transition frequency can be determined from the measured transition frequencies of a sufficient number of hyperfine components. However, in the experiments conducted so far, the number of measured hyperfine components has been too small, and no other independent experimental data on the relevant \hdplus{} hyperfine structure exist. To derive spin-averaged transition frequencies from limited sets of measured frequencies of hyperfine components, several methods have been developed which overcome the lack of experimental data by using the theoretically predicted hyperfine level structure~\cite{Alighanbari2020,Koelemeij2022,Korobov22}. In this work we adopt the approach of \cite{Koelemeij2022}, which is based on a least-squares optimization and takes into account correlations between theoretical hyperfine coefficients.

This article is organised as follows. The state-of-the-art \hdplus{} hyperfine theory used for the extraction of spin-averaged transition frequencies is reviewed in Section~\ref{sec-theory}, along with a 
discussion of correlated hyperfine theory uncertainties. 
Section~\ref{sec-exp} summarises the available experimental data (that include hyperfine shifts) related to three different rotational-vibrational transitions in \hdplus{}, followed by a global adjustment of the three spin-averaged rotational-vibrational transition frequencies in Section~\ref{sec-globaladj}. In Section~\ref{sec-discussion}, the adjusted spin-averaged transition frequencies and their covariances are subsequently compared to theoretical expressions for these spin-averaged transition frequencies that are functional expressions of several fundamental constants, including \Ar{e}, \Ar{p}, and the deuteron relative atomic mass, \Ar{d}. We use the corresponding observational equations in a second adjustment, aimed at determining the values of the fundamental constants. We consider various sets of input data and observational equations (including recent Penning-trap measurements of relative atomic masses), and compare values of mass ratios such as \mpme{} derived from each adjustment, in order to illustrate the potential impact of \hdplus{} on the precision of fundamental particle mass ratios. Conclusions are presented in Section~\ref{sec-concl}.

\section{Status of hyperfine structure theory} \label{sec-theory}

\subsection{Hyperfine Hamiltonian} \label{sec-hfs-ham}

The hyperfine structure of a rovibrational state of \hdplus{} can be described by the following effective spin Hamiltonian~\cite{Bakalov06}:
\begin{equation} \label{hdplus-heff}
\begin{split}
H_{\rm eff} &= E_1 (\mathbf{L}\!\cdot\!\mathbf{s}_{\rm e}) + E_2 (\mathbf{L}\!\cdot\!\mathbf{I}_{\rm p}) + E_3 (\mathbf{L}\!\cdot\!\mathbf{I}_{\rm d}) + E_4 (\mathbf{I}_{\rm p}\!\cdot\!\mathbf{s}_{\rm e}) \\
&\quad + E_5 (\mathbf{I}_{\rm d}\!\cdot\!\mathbf{s}_{\rm e})
+ E_6 \{ 2 \mathbf{L}^2 (\mathbf{I}_{\rm p}\!\cdot\!\mathbf{s}_{\rm e}) \!-\! 3 [(\mathbf{L}\!\cdot\!\mathbf{I}_{\rm p}) (\mathbf{L}\!\cdot\!\mathbf{s}_{\rm e}) \\
&\quad + (\mathbf{L}\!\cdot\!\mathbf{s}_{\rm e}) (\mathbf{L}\!\cdot\!\mathbf{I}_{\rm p})] \}
+ E_7 \{ 2 \mathbf{L}^2 (\mathbf{I}_{\rm d}\!\cdot\!\mathbf{s}_{\rm e}) \\
&\quad -3 [(\mathbf{L}\!\cdot\!\mathbf{I}_{\rm d}) (\mathbf{L}\!\cdot\!\mathbf{s}_{\rm e}) \!+\! (\mathbf{L}\!\cdot\!\mathbf{s}_{\rm e}) (\mathbf{L}\!\cdot\!\mathbf{I}_{\rm d})] \} \\
&\quad + E_8 \{ 2 \mathbf{L}^2 (\mathbf{I}_{\rm p}\!\cdot\!\mathbf{I}_{\rm d}) \!-\!3 [(\mathbf{L}\!\cdot\!\mathbf{I}_{\rm p}) (\mathbf{L}\!\cdot\!\mathbf{I}_{\rm d}) \!+\! (\mathbf{L}\!\cdot\!\mathbf{I}_{\rm p}) (\mathbf{L}\!\cdot\!\mathbf{I}_{\rm d})] \} \\
&\quad + E_9 \{ \mathbf{L}^2 \mathbf{I}_{\rm d}^2 \!-\! \frac{3}{2} (\mathbf{L}\!\cdot\!\mathbf{I}_{\rm d}) \!-\! 3 (\mathbf{L}\!\cdot\!\mathbf{I}_{\rm d})^2 \},
\end{split}
\end{equation}
where $\mathbf{I}_{\rm d}$, $\mathbf{I}_{\rm p}$, and $\mathbf{s}_{\rm e}$ are the spins of the deuteron, proton, and electron, respectively, and $\mathbf{L}$ is the total angular momentum excluding electron and nuclear spins (denoted by $\mathbf{N}$ in standard spectroscopic notation). Here and throughout the paper, spin and angular momentum operators are written in units of $\hbar$ and are thus dimensionless. The coefficients $E_k$, $k=1,2,\ldots,9$ are then energies and depend on the rovibrational state $(v,L)$, where $v$ stands for the vibrational quantum number. This Hamiltonian includes all spin-dependent interactions that appear at the leading order $m_{\rm e}c^2\alpha^4$, i.e. appear in the Breit-Pauli Hamiltonian, except for the proton-deuteron spin-spin contact interaction, which is negligibly small because of the strong Coulomb repulsion between the proton and deuteron. 

Additional spin-dependent interactions appear at higher orders ($m_{\rm e}c^2\alpha^6$ and above). The largest missing term in Equation~(\ref{hdplus-heff}) is a coupling between proton and deuteron spins (proportional to $\mathbf{I}_{\rm p}\!\cdot\!\mathbf{I}_{\rm d}$), mediated by the electron. The coupling coefficient of this interaction was shown to be on the order of 100~Hz~\cite{Korobov22}, which is much smaller than the current theoretical uncertainty of $E_4$ (see below). The effective spin Hamiltonian may thus be considered complete for our purposes.

The largest terms in Equation~(\ref{hdplus-heff}) are the electron-proton ($E_4/h \sim 900$~MHz in the vibrational ground state) and electron-deuteron ($E_5/h \sim 140$~MHz) spin-spin contact interactions. They were calculated with relative uncertainties below 1 parts per million (ppm)~\cite{Karr2020}. A key element to reach this accuracy is that the sum of nuclear structure and recoil corrections to $E_4$ ($E_5$), which we denote as 'nuclear correction' in the following, was determined from the difference between the experimental ground-state hyperfine splitting of the hydrogen (deuterium) atom and the corresponding nonrecoil, point-particle QED prediction, under the assumption that the nuclear correction is entirely described by a contact (delta-function) term. Uncertainties associated with nuclear corrections are thus suppressed, and the theoretical precision is limited by unevaluated higher-order QED terms: (i) nonrecoil corrections, dominated by the contribution of order $\alpha(Z\alpha^2) \ln(Z\alpha) E_F$~\cite{Layzer64,Zwanziger64}, and (ii) recoil corrections that deviate from the contact-term approximation, i.e. mainly the contribution of order $(Z\alpha)^2(m/M) E_F$~\cite{Bodwin88}. Here, $E_F$ stands for the leading Fermi contribution of order $m_{\rm e}c^2\alpha^4$.

The next largest terms are the spin-orbit ($E_1/h \sim 30$~MHz in the vibrational ground state) and spin-spin tensor interactions ($E_6/h \sim 8$~MHz, $E_7/h \sim 1$~MHz for $L=1$). They were computed with few-ppm uncertainties through calculation of higher-order corrections at orders $m_{\rm e}c^2\alpha^6$~\cite{Korobov20} and $m_{\rm e}c^2\alpha^7\ln(\alpha)$~\cite{Haidar22a,Haidar22b}. The uncertainty is dominated by unevaluated nonlogarithmic corrections at the $m_{\rm e}c^2\alpha^7$ order.

The remaining hyperfine coefficients are much smaller, ranging from $E_8/h \sim 3$~kHz to $E_2/h \sim 30$~kHz. They were calculated in the framework of the Breit-Pauli Hamiltonian accounting for the electron's anomalous magnetic moment~\cite{Bakalov06}. In general, the expected relative uncertainty of this approximation is on the order of $\alpha^2$, but in this case second-order contributions induced by the leading hyperfine interaction terms can give larger contributions. Due to this, the relative uncertainty of $E_2$, $E_3$, and $E_8$ is estimated to be $5\alpha^2$~\cite{Haidar22b}. In the case of the $E_9$ term that describes the effect of the deuteron quadrupole moment, second-order terms are too small to significantly influence the uncertainty. The relative uncertainty of $E_9$ is estimated to be 98~ppm~\cite{Haidar22b}, which accounts for both the theoretical uncertainty and the uncertainty of the deuteron quadrupole moment from~\cite{Puchalski20}.

State-of-the art values of all the hyperfine coefficients, for rovibrational levels involved in published high-precision measurements, are given in Table~\ref{table-hfs-coeff}.

In our least-squares adjustment of \hdplus{} experimental data (see Section~\ref{sec-globaladj}), theoretical uncertainties in the hyperfine coefficients $E_k(v,L)$ are taken into account by introducing additive corrections $\delta E_k^{\rm th} (v,L)$ which are treated as adjusted constants, following the approach of CODATA~\cite{Tiesinga2021}. For each coefficient, an input datum $\delta E_k(v,L)$ is included with a zero value and an uncertainty equal to that given in Table~\ref{table-hfs-coeff}, and an observational equation $\delta E_k (v,L) \doteq \delta E_k^{\rm th} (v,L)$ is added to the adjustment. 

Beyond the uncertainties themselves, it is also necessary to estimate the correlation coefficients among the corrections $\delta E_k (v,L)$ due to common sources of uncertainty. As shown in~\cite{Koelemeij2022}, the assumed level of correlation can significantly influence the spin-averaged frequencies of rovibrational transitions determined from an incomplete set of measurements of their hyperfine components. Correlation coefficients are defined as follows:
\begin{equation}
r(\delta E_k(v,L),\delta E_l(v',L')) = \frac{u (\delta E_k(v,L), \delta E_l(v',L'))}{u (\delta E_k(v,L)) \, u(\delta E_l(v',L'))} \,,  
\label{eq:corrcoeffdef}\end{equation}
where $u(X,Y)$ is the covariance between quantities $X$ and $Y$, and $u(X) = \sqrt{u(X,X)}$ is the uncertainty of $X$.

Here, we make assumptions similar to those presented in Appendix C of~\cite{Haidar22b}:
\begin{itemize}
	\item Perfect positive correlations are assumed between $\delta E_4(v,L)$ and $\delta E_4(v',L')$, as well as between $\delta E_5(v,L)$ and $\delta E_5(v',L')$.
	\item For the other hyperfine coefficients ($k \neq 4,5$), correlation coefficients between $\delta E_k(v,L)$ and $\delta E_k(v',L')$ are expected to be positive, but are difficult to estimate. In our analysis, we thus vary these coefficients between 0 and 1 to investigate their influence on the spin-averaged frequencies, and additional uncertainties associated with these correlations.
	\item Perfect positive correlations are assumed between $\delta E_2(v,L)$ and $\delta E_3(v,L)$, as well as between $\delta E_6(v,L)$ and $\delta E_7(v,L)$. This stems from the fact that uncalculated QED corrections in both coefficients are of the same nature.
	\item The correlation coefficient between $\delta E_4(v,L)$ and $\delta E_5(v,L)$ is estimated~\cite{Haidar22b} to be $r(\delta E_4(v,L), \delta E_5(v,L)) = 0.4016$.
	\item No other correlations are assumed between $\delta E_k(v,L)$ and $\delta E_l(v,L)$.
	\item No correlations are assumed between $\delta E_k(v,L)$ and $\delta E_l(v',L')$ with $k \neq l$ and $(v,L) \neq (v',L')$, except for those that are implied by the above hypotheses, e.g.: $r(\delta E_4 (v,L), \delta E_5(v',L')) = r(\delta E_4(v,L), \delta E_5(v,L)) = 0.4016$.
\end{itemize}
Except for the unknown correlation between $\delta E_k(v,L)$ and $\delta E_k(v',L')$ ($k \neq 4,5$), we do not consider the effect of uncertainties in other correlation coefficients. Indeed, although these uncertainties are not strictly zero, they are estimated to be too small to have any significant impact on the spin-averaged frequencies. 

\begin{table*}[h!]
	\scalebox{0.65}{
	\hspace*{-2cm}	\begin{tabular}{|c|S[table-format=5.8]|S[table-format=4.6]|S[table-format=3.6]|S[table-format=6.8]|S[table-format=6.7]|S[table-format=4.7]|S[table-format=4.6]|S[table-format=3.6]|S[table-format=1.6]|}
			\hline
			{$(v,L)$} & {$E_1/h$} & {$E_2/h$} & {$E_3/h$} & {$E_4/h$} & {$E_5/h$} & {$E_6/h$} & {$E_7/h$} & {$E_8/h$} & {$E_9/h$} \\
			\hline
			$(0,0)$ & & & & 925394.159(860) & 142287.556(84) & & & & \\
			$(0,1)$ & 31 985.417(116) & -31.345(8) & -4.809(1) & 924 567.718(859) & 142 160.670(84) & 8 611.299(18) & 1 321.796(3) & -3.057(1) & 5.660(1) \\
			$(1,1)$ & 30 280.736(109) & -30.463(8) & -4.664(1) & 903 366.501(839) & 138 910.266(82) & 8 136.858(17) & 1 248.963(3) & -2.945(1) & 5.653(1) \\
			$(0,3)$ & 31 628.097(114) & -30.832(8) & -4.733(1) & 920 479.981(855) & 141 533.075(83) & 948.542(2) & 145.597 & -0.335 & 0.613 \\
			$(9,3)$ & 18 270.853(62) & -21.304(6) & -3.225(1) & 775 706.122(721) & 119 431.933(73) & 538.999(1) & 82.726 & -0.219 & 0.501 \\
			\hline
	\end{tabular}}
	\caption{Theoretical hyperfine coefficients of rovibrational states of \hdplus{} involved in high-precision measurements, in kHz. Missing numbers in the first line imply a zero value. The values of $E_4/h$, $E_5/h$ ($E_1/h$, $E_6/h$, $E_7/h$) were calculated in Ref.~\cite{Karr2020} (Ref.~\cite{Haidar22b}). Those of $E_2/h$, $E_3/h$, $E_8/h$, and $E_9/h$ are taken from~\cite{Bakalov06}. The value of $E_9/h$ has been updated using the latest determination of the deuteron's quadrupole moment~\cite{Puchalski20}. \label{table-hfs-coeff}}
\end{table*}

\subsection{Hyperfine splitting}

The hyperfine structure of a rovibrational state $(v,L)$ is obtained by diagonalizing the effective spin Hamiltonian~(\ref{hdplus-heff}). Given the relative sizes of the $E_k$, the natural coupling scheme of angular momenta is~\cite{Bakalov06}
\begin{equation}
\mathbf{F} = \mathbf{I}_{\rm p} + \mathbf{s}_{\rm e}, \quad \mathbf{S} = \mathbf{F} + \mathbf{I}_{\rm d}, \quad \mathbf{J} = \mathbf{L} + \mathbf{S}.
\end{equation}
The matrix elements of the effective spin Hamiltonian in the basis of coupled states $| v L F S J \rangle$ are calculated using standard angular algebra procedures. $J$ is an exact quantum number, whereas $F$ and $S$ are only approximate quantum numbers. A spin eigenstate is denoted by $| v L \tilde{F} \tilde{S} J \rangle$, which is the state having the largest weight in its decomposition over the basis states. In the following, we omit the $\tilde{}$ symbols for simplicity.

Since the matrix of $H_{\rm eff}$ is block diagonal with blocks of size $4\times4$ at most, its eigenvalues $E_{\rm hfs} (v,L,F,S,J)$ can be expressed analytically as a function of the hyperfine interaction coefficients $E_k$. These expressions are then linearised around the theoretical values of the coefficients using the sensitivity coefficients
\begin{equation}
\gamma_k (v,L,F,S,J) = \frac{\partial E_{\rm hfs} (v,L,F,S,J)} {\partial E_k}.
\end{equation}
The theoretical hyperfine shifts $E_{\rm hfs}$ and sensitivity coefficients $\gamma_k$ of all the levels involved in the measurements are given in Table~\ref{table-sensitHFS}. 
\begin{table*}[h!]
    \scalebox{0.7}{
	\hspace*{-2cm}\begin{tabular}{|c|c|S[table-format=7.7]|*{6}{S[table-format=3.4]|}*{2}{S[table-format=4.4]|}S[table-format=3.4]|}
		\hline
		{$(v,L)$} & {$(F,S,J)$} & {$E_{\rm hfs}/h$ (kHz)} & {$\gamma_1$} & {$\gamma_2$} & {$\gamma_3$} & {$\gamma_4$} & {$\gamma_5$} & {$\gamma_6$} & {$\gamma_7$} & {$\gamma_8$} & {$\gamma_9$} \\
		\hline
		\multirow{4}{*}{$(0,0)$}
		& $(0,1,1)$ & -705 735.655(639) & & & & -0.7367 & -0.1688 & & & & \\
		& $(1,0,0)$ &   89 060.984(197) & & & &  0.2500 & -1.0000 & & & & \\
		& $(1,1,1)$ &  171 894.797(194) & & & &  0.2367 & -0.3312 & & & & \\
		& $(1,2,2)$ &  302 492.318(235) & & & &  0.2500 &  0.5000 & & & & \\
		\hline
		\multirow{6}{*}{$(0,1)$}
		& $(0,1,2)$ & -707 870.694(637) & -0.1070 &  0.1110 &  0.9953 & -0.7338 & -0.1845 &  0.0100 &
		0.1434 & -0.1572 & -0.4930 \\
		& $(1,0,1)$ &   79 986.308(204) & -0.4286 & -0.3861 &  0.6539 &  0.2494 & -0.9084 & -1.0279 &
		0.8343 &  0.7206 & -0.5019 \\
		& $(1,1,2)$ &  183 682.795(196) &  0.3319 &  0.1539 &  0.5142 &  0.2341 & -0.3152 &  0.2555 &
		-0.5821 & -0.4019 &  0.2287 \\
		& $(1,2,1)$ &  269 283.177(241) & -0.5693 & -0.5592 & -1.7183 &  0.2500 &  0.4251 &  0.0392 &
		-3.3471 & -3.2839 & -2.9444 \\
		& $(1,2,3)$ &  312 567.273(242) &  0.5000 &  0.5000 &  1.0000 &  0.2500 &  0.5000 & -0.5000 &
		-1.0000 & -1.0000 & -0.5000 \\
		& $(1,2,2)$ &  314 228.466(238) & -0.2249 & -0.2649 & -0.5096 &  0.2497 &  0.4997 &  1.7345 &
		3.4387 &  3.5591 &  1.7643 \\
		\hline
		\multirow{2}{*}{$(1,1)$}
		& $(1,2,1)$ &  263 806.300(235) & -0.5753 & -0.5654 & -1.7152 &  0.2500 &  0.4296 & -0.0105 &
		-3.3687 & -3.3063 & -2.9087 \\
		& $(1,2,3)$ &  305 099.957(236) &  0.5000 &  0.5000 &  1.0000 &  0.2500 &  0.5000 & -0.5000 &
		-1.0000 & -1.0000 & -0.5000 \\
		\hline
		\multirow{2}{*}{$(0,3)$}
		& $(0,1,4)$ & -711 832.144(631) & -0.4142 &  0.4242 &  2.9867 & -0.7267 & -0.2156 &  0.2750 &
		3.0796 & -3.3289 & -7.4000 \\
		& $(1,2,5)$ &  338 970.635(290) &  1.5000 &  1.5000 &  3.0000 &  0.2500 &  0.5000 & -7.5000 &
		-15.0000 & -15.0000 & -7.5000 \\
		\hline
		\multirow{2}{*}{$(9,3)$}
		& $(0,1,4)$ & -596 833.613(533) & -0.3590 &  0.3727 &  2.9847 & -0.7303 & -0.2017 &  0.1934 &
		2.4743 & -2.7221 & -7.3853 \\
		& $(1,2,5)$ &  275 723.281(219) &  1.5000 &  1.5000 &  3.0000 &  0.2500 &  0.5000 & -7.5000 &
		-15.0000 & -15.0000 & -7.5000 \\
		\hline
	\end{tabular}}
	\caption{Theoretical hyperfine shifts (column 3) and sensitivity coefficients (columns 4-12) for all the levels involved in high-precision measurements. Missing numbers for the $(v,L)=(0,0)$ state imply a zero value. The uncertainties of the hyperfine shifts are calculated assuming no correlation between $\delta E_k(v,L)$ and $\delta E_k(v',L')$ for $k \neq 4,5$, which provides an upper bound of the uncertainty. \label{table-sensitHFS}}
\end{table*}
\section{Available experimental data} \label{sec-exp}
At the time of writing, ppt-range frequency measurements of three rotational-vibrational transitions in \hdplus{} had been reported in peer-reviewed publications. These are the \vRot{} rotational transition at 1.31~THz~\cite{Alighanbari2020}, the \vVib{} rovibrational transition at 58.6~THz~\cite{Kortunov2021}, and the \vTP{} vibrational overtone at 415~THz~\cite{Patra2020}. All frequency measurements were made on \hdplus{} ions stored in an rf trap, and sympathetically cooled to millikelvin temperatures via laser-cooled Be$^+$ ions stored in the same trap. For further details on the experimental setups and the justification of the achieved measurement uncertainties we refer to the original publications.

Table~\ref{table-expinput} provides an overview of the frequencies (and their uncertainties) of the 10 measured hyperfine components belonging to the three rotational-vibrational transitions of interest. The measured frequencies are denoted by $f^{\rm exp} [(v,L,a) \to (v',L',a')]$, where $a=(F,S,J)$ and $a'=(F',S',J')$ are the initial and final hyperfine states.

In all cases the experimental uncertainties receive appreciable contributions from statistical as well as systematic sources of uncertainty. In the uncertainty evaluation of the \vRot{} and \vVib{} transitions, no mention is made of possible correlated systematic errors~\cite{Alighanbari2020,Kortunov2021}. In the case of the \vTP{} transition, a correlation between two negligibly small (sub-0.01-kHz) uncertainties in the estimate of the Zeeman shift was identified~\cite{Patra2020}. In what follows we assume that the uncertainties of all experimental transition frequencies are uncorrelated.
\begin{table*}
	\scalebox{0.88}{
        \begin{tabular}{|c|c|c|c|r|c|c|} 
		\hline
		$(v\!,\!L)\!\to\! (v'\!,\!L')$ & Label & $(F\!,\!S\!,\!J)$ & $(F'\!,\!S'\!,\!J')$ & \multicolumn{1}{|c|}{$f^{\rm exp} [(v\!,\!L\!,\!a) \!\to\! (v'\!,\!L'\!,\!a')]$ (kHz)} & Alt. label & Ref. \\
		\hline
		\multirow{6}{*}{$(0,0)\rightarrow (0,1)$}
		& A1 & (1,2,2) & (1,2,1) & \text{1,314,892,544.276(40)} & 12 & \multirow{6}{*}{\cite{Alighanbari2020}}\\
		& A2 & (1,0,0) & (1,0,1) & \text{1,314,916,678.487(64)} & 14 &  \\
		& A3 & (0,1,1) & (0,1,2) & \text{1,314,923,618.028(17)} & 16 & \\
		& A4 & (1,2,2) & (1,2,3) & \text{1,314,935,827.695(37)} & 19 & \\
		& A5 & (1,2,2) & (1,2,2) & \text{1,314,937,488.614(60)} & 20 & \\
		& A6 & (1,1,1) & (1,1,2) & \text{1,314,937,540.762(46)} & 21 &  \\
		\hline
		\multirow{2}{*}{$(0,0)\rightarrow (1,1)$}
		& A7 & (1,2,2) & (1,2,1) & \text{58,605,013,478.03(19)} & 12 & \multirow{2}{*}{\cite{Kortunov2021}}\\
		& A8 & (1,2,2) & (1,2,3) & \text{58,605,054,772.08(26)} & 16 & \\
		\hline
		\multirow{2}{*}{$(0,3)\rightarrow (9,3)$}
		& A9 & (0,1,4) & (0,1,4) & \text{415,265,040,503.57(59)} & $F=0$ & \multirow{2}{*}{\cite{Patra2020}} \\
		& A10 & (1,2,5) & (1,2,5) & \text{415,264,862,249.16(66)} & $F=1$ & \\
		\hline
	\end{tabular}}
	\caption{Experimentally determined transition frequencies, $f^{\rm exp} [(v,L,a) \to (v',L',a')]$ (where $a=(F,S,J)$ and $a'=(F',S',J')$), of various hyperfine components used in the adjustment of the spin-averaged transition frequencies (Section \ref{sec-globaladj}). The alternate labels for the hyperfine components in column 6 are those used in the original references (column 7). \label{table-expinput}}
\end{table*}
\section{Global adjustment of spin-averaged transition frequencies} \label{sec-globaladj}

\subsection{Adjustment details}
\label{sec-globaladjA}

The spin-averaged transition frequencies are determined from a least-squares adjustment of the available theoretical and experimental data presented in Secs.~\ref{sec-theory} and ~\ref{sec-exp}, respectively. We follow the fitting procedure described in Appendix E of~\cite{Mohr2000}.

All experimental data listed in Table~\ref{table-expinput} are included as input data for the adjustment. The associated ``observational equations'' (see~\cite{Mohr2000} for details) read
\begin{equation}
\begin{split}
&\,f^{\rm exp} [(v,L,a) \to (v',L',a')] \\
&\doteq f_{\rm SA} [(v,L) \!\to\! (v',L')] \!+\! \left( E_{\rm hfs} (v'\!,\!L'\!,\!a') \!-\! E_{\rm hfs} (v\!,\!L\!,\!a) \right)/h \\
& \quad + \frac{1}{h}\sum_{k=1}^9 ( \gamma_k (v'\!,\!L'\!,\!a') \, \delta E_k^{\rm th} (v'\!,\!L') \!-\! \gamma_k (v\!,\!L\!,\!a) \, \delta E_k^{\rm th} (v\!,\!L) ),
\end{split}
\end{equation}
where the dotted equality sign means that the left and right hand sides are not equal in general (since the set of equations is overdetermined) but should agree within estimated uncertainties. Here, $f_{\rm SA} [(v,L) \to (v',L')]$ are the spin-averaged frequencies of the rovibrational transitions. Moreover, there are observational equations associated with the additive corrections $\delta E_k^{\rm th} (v,L)$ to theoretical hyperfine coefficients, as explained in Section~\ref{sec-hfs-ham}:
\begin{equation}
\delta E_k (v,L) \doteq \delta E_k^{\rm th} (v,L).    
\end{equation}
The adjusted constants are $f_{\rm SA} [(v,L) \!\to\! (v',L')]$ and $\delta E_k^{\rm th} (v,L)$. However, some of the $\delta E_k (v,L)$ are redundant because certain pairs of coefficients are perfectly correlated with each other (see Section~\ref{sec-hfs-ham}). This leads to a singular covariance matrix, which cannot be inverted as required for the least-squares adjustment. To solve this, redundant variables are eliminated using the relationships:
\begin{subequations}
	\begin{gather}
	\delta E_k (v,L) = \delta E_k (0,0) \frac{u_k (v,L)}{u_k(0,0)}, \quad k=4,5, \\
	\delta E_l (v,L) = \delta E_k (v,L) \frac{u_l(v,L)}{u_k(v,L)}, \; (k,l)=(2,3) \; \mbox{or} \; (6,7),
	\end{gather}
\end{subequations}
where $u_k (v,L)$ is the uncertainty of $\delta E_k (v,L)$. There are a total of $N = 32$ input data: 10 experimental frequencies (Table~\ref{table-expinput}), and 22 additive energy corrections (Table~\ref{table-theorinput}). The total number of adjusted constants is $M = 25$: 3 spin-averaged transition frequencies and the 22 energy corrections.

\begin{table*}[h!]
	\begin{tabular}{@{\hspace{2mm}}c@{\hspace{2mm}}c@{\hspace{2mm}}S[table-format=1.7]@{\hspace{8mm}}c@{\hspace{2mm}}c@{\hspace{2mm}}S[table-format=1.8]@{\hspace{2mm}}}
		\hline
		& Input datum       & {Value (kHz)} &    & Input datum        & {Value (kHz)} \\
		B1  & $\delta E_4 (0,0)/h$ & 0.000(860)  & B12 & $\delta E_9 (1,1)/h$ & 0.00000(55)  \\
		B2  & $\delta E_5 (0,0)/h$ & 0.000(84)   & B13 & $\delta E_1 (0,3)/h$ & 0.000(114)   \\
		B3  & $\delta E_1 (0,1)/h$ & 0.000(116)  & B14 & $\delta E_2 (0,3)/h$ & 0.0000(82)   \\
		B4  & $\delta E_2 (0,1)/h$ & 0.0000(83)  & B15 & $\delta E_6 (0,3)/h$ & 0.0000(20)   \\
		B5  & $\delta E_6 (0,1)/h$ & 0.000(18)   & B16 & $\delta E_8 (0,3)/h$ & 0.000000(89) \\
		B6  & $\delta E_8 (0,1)/h$ & 0.00000(81) & B17 & $\delta E_9 (0,3)/h$ & 0.000000(60) \\
		B7  & $\delta E_9 (0,1)/h$ & 0.00000(55) & B18 & $\delta E_1 (9,3)/h$ & 0.000(62)    \\
		B8  & $\delta E_1 (1,1)/h$ & 0.000(109)  & B19 & $\delta E_2 (9,3)/h$ & 0.0000(57)  \\
		B9  & $\delta E_2 (1,1)/h$ & 0.0000(81)  & B20 & $\delta E_6 (9,3)/h$ & 0.0000(12)   \\
		B10 & $\delta E_6 (1,1)/h$ & 0.000(17)   & B21 & $\delta E_8 (9,3)/h$ & 0.000000(58) \\
		B11 & $\delta E_8 (1,1)/h$ & 0.00000(78) & B22 & $\delta E_9 (9,3)/h$ & 0.000000(49) \\
		\hline
	\end{tabular}
	\caption{Input data for the additive energy corrections to account for the theoretical uncertainties of hyperfine interaction coefficients. \label{table-theorinput}}
\end{table*}

\subsection{Effect of unknown  hyperfine correlation coefficients}\label{sec-globaladjB}
As discussed in Section~\ref{sec-hfs-ham}, the level of correlation between some of the theoretical hyperfine coefficients (namely, $\delta E_k(v,L)$ and $\delta E_k(v',L')$ with $k \neq 4,5)$ is not well known. We thus varied their correlation coefficients between 0 and 1 to study their influence on spin-averaged frequencies. These frequencies were found to be mostly sensitive to the correlation between $\delta E_1(v,L)$ and $\delta E_1(v',L')$, with a smaller effect from the correlation between $\delta E_6(v,L)$ and $\delta E_6(v',L')$ (which themselves are fully correlated with $\delta E_7(v,L)$ and $\delta E_7(v',L')$, respectively). Correlations involving $E_2$, $E_3$, $E_8$, and $E_9$ have a negligible impact. In order to estimate the additional shift, $f_{\rm corr} (v,L,v',L')$, and its uncertainty, $\delta f_{\rm corr} (v,L,v',L')$, caused by the unknown correlation coefficients, we carried out a Monte-Carlo simulation similar to \cite{Koelemeij2022}, as further detailed in Appendix~\ref{app-monte-carlo}. We find that the mean shift of the simulated spin-averaged transition frequencies is well accounted for when carrying out a single adjustment using correlation coefficients of 0.5. Furthermore, we determine the full widths of the simulated frequency distributions to be 2.09~Hz, 144~Hz, and 237~Hz for the rotational, $v=0 \to 1$, and $v=0 \to 9$ transitions, respectively. We conservatively define $\delta f_{\rm corr} (v,L,v',L')$ to be equal to the full width divided by two. As will be shown below, these correlation-induced uncertainties are considerably smaller than the frequency uncertainties resulting from the adjustment itself. 

The question is whether the uncertainties $\delta f_{\rm corr} (v,L,v',L')$ are correlated (which would introduce correlations between different transition frequencies $f_{\rm SA} [(v,L) \!\to\! (v',L')]$ in addition to the correlations that may result from the adjustment itself). We empirically find either strong correlations (i.e. $r(\delta f_{\rm corr} (v,L,v',L'),\delta f_{\rm corr} (v'',L'',v''',L''')) = +1$) or strong anticorrelations (i.e. $r(\delta f_{\rm corr} (v,L,v',L'),\delta f_{\rm corr} (v'',L'',v''',L'''))=-1$); see Appendix~\ref{app-monte-carlo}. We can thus construct a 'correlation-induced' covariance matrix from the correlation-induced uncertainties defined above, in combination with the empirically found correlation coefficients.     
\subsection{Results and discussion}\label{sec-globaladjC}
In Table~\ref{table-fSA}, we compare our derived spin-averaged frequencies with those in the literature. These literature values are based on the same experimental data as used in our analysis; literature values based on specific subsets of the experimental data (such as the value presented in~\cite{Korobov22}) are not considered here. 
The rotational transition is shifted by about four standard deviations with respect to the literature value of~\cite{Alighanbari2020}. This is mainly due to the fact that we make different assumptions regarding correlations between theoretical hyperfine coefficients, especially those between $E_4(0,0)$ and $E_4(0,1)$. As justified in Ref.~\cite{Haidar22b}, we assume perfect positive correlation, whereas zero correlation was assumed in the initial analysis~\cite{Alighanbari2020}. The sensitivity of the rotational transition on the assumed level of correlation was already observed in~\cite{Koelemeij2022} (see Fig~4 (b) in that reference). For the other transitions, the frequencies determined in this work are well within the error bars of the previous determinations.

As can be seen in Table~\ref{table-fSA}, the improvement of hyperfine structure theory~\cite{Haidar22b} has allowed the reduction of the uncertainties of the $v=0 \to 1$ and $v=0 \to 9$ transition frequencies with respect to their initial determinations~\cite{Patra2020,Kortunov2021}. For the rotational transition, the precision has only slightly improved, because in this case 6 spin components have been measured, making the extracted frequency much less sensitive to theoretical uncertainties.

\begin{figure}[h!]
	\centering
	\includegraphics[width=0.7\columnwidth]{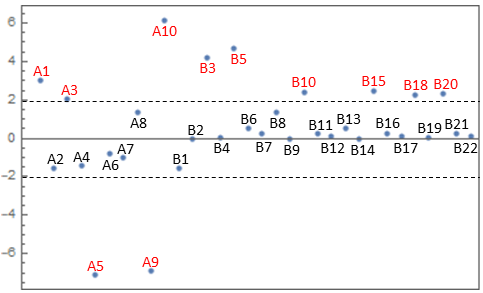}
	\caption{
		Normalised residuals of the 32 input data for the adjustment of spin-averaged transition frequencies. Labels A{\it i} and B{\it i} follow those defined in Tables~\ref{table-expinput} and \ref{table-theorinput}.\label{fig-residuals}
	}
\end{figure}

The normalised residuals of the adjustment (assuming a value of 0.5 for the unknown correlation coefficients) are shown in Figure~\ref{fig-residuals}. They reveal substantial discrepancies, with normalised residuals larger than two, between the experimental data and the theoretical hyperfine structure:
\begin{itemize}
	\item the hyperfine components A1, A3, and A5 of the rotational transition (with labels A{\it i} defined in Table~\ref{table-expinput}) have discrepancies of $3.0\sigma$ (121~Hz), $2.1\sigma$ (35~Hz), and $-7.1\sigma$ ($-427$~Hz), respectively;
	\item both components of the $v=0 \to 9$ transition, A9 and A10, exhibit discrepancies of $-6.9\sigma$ ($-4.5$~kHz) and $6.1\sigma$ (3.6~kHz), respectively;
	\item several theoretical spin coefficients, in particular $E_1 (v=0,L=1)$ and $E_6 (v=0,L=1)$, with deviations of $4.2\sigma$ ($h\times484$~Hz) and $4.7\sigma$ ($h\times85$~Hz), respectively. Deviations slightly above 2$\sigma$ are also found in $E_6 (v=1,L=1)$, $E_6 (v=0,L=3)$, $E_1 (v=9,L=3)$, and $E_6 (v=9,L=3)$.
\end{itemize}
Discrepancies in the hyperfine structure of the $L=0 \to 1$ and $v=0\to 9$ transitions had already been discussed in~\cite{Karr2020,Haidar22b}. Their origin is presently unknown; they could be due to a problem in the theory, in the experiments, or both. It is furthermore worth mentioning that experimental data on the hyperfine structure of H$_2^+$ are in good agreement with theoretical predictions; see \cite{Haidar22b} and references therein.

We treat these deviations following the procedure used in the CODATA adjustments (see, e.g.,~\cite{Tiesinga2021}). A multiplicative expansion factor is applied to the initially assigned uncertainties of all (experimental and theoretical) input data, such that the absolute values of all normalised residuals are smaller than two. The required expansion factor is $\eta=3.56$. As a result, the uncertainties of the spin-averaged frequencies are multiplied by the same factor (third line in Table~\ref{table-fSA}). The expanded uncertainties are one order of magnitude larger than the frequency uncertainties caused by the unknown correlation coefficients.

The spin-averaged frequencies obtained by the least-squares adjustment have small but nonzero correlations between them due to correlations between $\delta E_k$ in different rovibrational levels. The magnitude of the corresponding correlation coefficients is less than 0.01, and their values (and in some case also their signs) depend on the choice of hyperfine correlation coefficients (Section~\ref{sec-hfs-ham}). Therefore, the correlations between the three adjusted spin-averaged transition frequencies have an uncertainty due to the unknown hyperfine theory correlations. In Section~\ref{sec-discussion} we show that these correlations have a negligible impact on the fundamental constants that may be derived from the \hdplus{} data, and we do not investigate this particular source of uncertainty further here. However, for completeness we do combine the covariance matrix resulting from the adjustment with the correlation-induced covariance matrix that was constructed as described in Section~\ref{sec-globaladjB}. The resulting (combined) uncertainties of the spin-averaged transition frequencies are listed in the last line of Table~\ref{table-fSA}, and their (combined) correlation coefficients are presented in Table~\ref{table-correl}.

\begin{table*}[h!]
    \scalebox{0.85}{
\hspace*{-1cm}	\begin{tabular}{|@{\hspace{2mm}}l@{\hspace{2mm}}|@{\hspace{2mm}}c@{\hspace{2mm}}|@{\hspace{2mm}}c@{\hspace{2mm}}|@{\hspace{2mm}}c@{\hspace{2mm}}|}
		\hline
		& $(0,0) \to (0,1)$ & $(0,0) \to (1,1)$ & $(0,3) \to (9,3)$ \\
		\hline
		$f_{\rm SA}$ (previous) & 1 314 925 752.910(17) & 58 605 052 164.24(86) & 415 264 925 500.5(1.2) \\
		\hline
		$f_{\rm SA}$ ($\eta=1$)  & 1 314 925 752.978(14) & 58 605 052 164.14(16) & 415 264 925 501.3(0.4) \\
		\hline
		$f_{\rm SA}$ ($\eta=3.56$) & 1 314 925 752.978(48) & 58 605 052 164.14(55) & 415 264 925 501.3(1.6) \\ 
		\hline
		$f_{\rm SA}$ ($\eta=3.56$ and corr. ind. unc.) & 1 314 925 752.978(48) & 58 605 052 164.14(56) & 415 264 925 501.3(1.6)               \\ 
		\hline
	\end{tabular}}
	\caption{Spin-averaged transition frequencies $f_{\rm SA}$ (in kHz). Previous determinations from the original publications~\cite{Alighanbari2020,Patra2020,Kortunov2021} are given in the first line. The results of the least-squares adjustment performed in the present work are given in line 2. These results were obtained with an expansion factor $\eta=1$ (i.e. no expansion factor applied). The spin-averaged frequencies in line 3 are obtained by applying an expansion factor $\eta=3.56$ to the uncertainties of all (experimental and theoretical) input data. Finally, line 4 shows the frequencies for $\eta=3.56$ and including the hyperfine-correlation-induced uncertainty; see text for details. The frequencies on line 4 are our recommended values. \label{table-fSA}}
\end{table*}

\section{Implications for the determination of particle masses} \label{sec-discussion}

Using the spin-averaged frequencies determined in this work, we can now assess the potential contribution of \hdplus{} spectroscopy to the determination of fundamental constants, in particular the relative atomic masses \Ar{e}, \Ar{p}, \Ar{d}, and related mass ratios. To do this, we use a least-squares adjustment following the same procedure as in the previous section, and compare three different scenarios:
\begin{itemize}
	\item Adjustment 1 only includes Penning trap measurements related to the electron, proton, and deuteron relative atomic masses;
	\item Adjustment 2 only includes \hdplus{} spectroscopic data and values of fundamental constants (other than atomic masses) involved in the theoretical energy levels of \hdplus{};
	\item Adjustment 3 includes all input data of Adjustments 1 and 2.
\end{itemize}

Theoretical rovibrational transition frequencies of \hdplus{} depend on~\cite{Karr2016}: the Rydberg constant ($R_{\infty}$), nucleus-to-electron mass ratios $\lambda_{\rm p} \equiv m_{\rm p}/m_{\rm e}$ and $\lambda_{\rm d} \equiv m_{\rm d}/m_{\rm e}$, the charge radii of the proton $r_{\rm p}$ and deuteron $r_{\rm d}$, and the fine-structure constant $\alpha$. The dependence on $\alpha$ is very weak because it only enters in the relativistic and QED corrections, and can be safely neglected without loss of precision. As noted in previous works, e.g.,~\cite{Korobov2017,Alighanbari2020}, the mass dependence can to a good approximation be reduced to a dependence on a single parameter $\lambda_{\rm pd} \equiv \mu_{\rm pd}/m_{\rm e}$, where $\mu_{\rm pd}$ is the proton-deuteron reduced mass. This is exactly true in the adiabatic approximation, but, due to nonadiabatic effects, there is actually a much weaker dependence on a second mass ratio, for example $\lambda_{\rm d}$. In our Adjustment 2, we neglect this weak dependence, allowing $\lambda_{\rm pd}$ to be determined from \hdplus{} spectroscopic data alone, as done in~\cite{Alighanbari2020,Kortunov2021}. Although $\lambda_{\rm pd}$ is not an adjustable parameter in Adjustment 1, it is possible to infer a value of $\lambda_{\rm pd}$ from the adjusted values of \Ar{e}, \Ar{p}, and \Ar{d}. The relative impact of \hdplus{} spectroscopy and Penning-trap measurements on future values of particle mass ratios can then be assessed by comparing the determinations of $\lambda_{\rm pd}$ (and their uncertainties) obtained from Adjustments 1 and 2. To further illustrate the effect of including \hdplus{} data, we also compare the values of $\lambda_{\rm p}$ (=\, \mpme{)} inferred from Adjustments 1 and 3.

In the perspective of a global adjustment of fundamental constants, it is preferable to take the dependence of the theoretical \hdplus{} transition frequencies on both parameters $\lambda_{\rm pd}$, $\lambda_{\rm d}$ into account in order to avoid any unnecessary loss of precision. Furthermore, for the global Adjustment 3, we have found it more convenient to parametrise the mass dependence of \hdplus{} transition frequencies in terms of the quantities that are treated as adjusted parameters in the CODATA adjustments~\cite{Tiesinga2021}, i.e., the relative atomic masses \Ar{e}, \Ar{p}, and \Ar{d}.

\subsection{Adjustment details}

Let us briefly describe Adjustment 1. It includes the 2018 CODATA value of electron's relative atomic mass, for which no new measurement has been reported since then. For the proton relative atomic mass, we use the value of $A_{\rm r}(^1{\rm H})$ from the latest (2020) Atomic Mass Evaluation (AME)~\cite{Wang:2021xhn,Huang:2021nwk}, corrected for the electron mass and for the theoretical binding energy. This value takes into account recent high-precision Penning trap data~\cite{Rau2020,Fink2020}, except for a very precise measurement of the deuteron-to-proton mass ratio, \mdmp{}, performed in 2021~\cite{Fink2021}. We thus also include the latter, together with the value of the deuteron mass deduced from the 2020 AME value of $A_{\rm r}(^2{\rm H})$ (again corrected for the electron mass and the binding energy), improving the proton  mass determination. Adjustment 1 thus comprises $N = 4$ input data and $M = 3$ adjusted parameters, the relative atomic masses of the electron, proton and deuteron.

The other adjustments include \hdplus{} spectroscopic data. Similarly to the hyperfine structure, theoretical uncertainties of spin-averaged transition frequencies are accounted for by introducing additive corrections $\delta f^{\rm th} (v,L,v',L')$ to the theoretical frequencies. They are treated as adjusted constants, and input data with a zero value and an uncertainty equal to the theoretical uncertainty of $f_{\rm SA}^{\rm th} [(v,L)\to (v',L')]$ are included (B1-B3 in Table~\ref{table-fc-input}). Their correlation coefficients (listed in Table~\ref{table-correl}) are estimated using the results of~\cite{Korobov2021}. Correlation coefficients are found to be close to 1, because the uncertainties are dominated by unevaluated QED terms, whereas numerical uncertainties are negligibly small.

In Adjustment 2, each of the three spin-averaged frequencies determined in the previous section (inputs A1-A3 in Table~\ref{table-fc-input}) is associated with an observational equation of the form
\begin{equation} \label{obs-eq-adj2}
\begin{split}
&\,f_{\rm SA} [(v,L)\to (v',L')] \\
&\doteq f_{\rm SA}^{\rm th} [(v,L)\to (v',L')] + \delta f^{\rm th} (v,L,v',L') \\
&\quad + \beta_{R_{\infty}} c(R_{\infty} \!-\! R_{\infty 0}) + \eta_{r_{\rm p}} (r_{\rm p}^2 \!-\! r_{{\rm p}0}^2) + \eta_{r_{\rm d}} (r_{\rm d}^2 \!-\! r_{{\rm d}0}^2) \\
&\quad + \beta_{\lambda_{\rm pd}} (\lambda_{\rm pd} \!-\! \lambda_{\rm pd0}) + \beta_{\lambda_{\rm d}} (\lambda_{\rm d} \!-\! \lambda_{\rm d0}).
\end{split}
\end{equation}
Here, $f_{\rm SA}^{\rm th} [(v,L)\to (v',L')]$ is the reference value of the theoretical transition frequency~\cite{Korobov2021}, calculated with 2018 CODATA values of fundamental constants. The reference (2018 CODATA) values of $\ArM{e}$, $R_{\infty}$, $r_{\rm p}$, and $r_{\rm d}$ are given in Table~\ref{table-fc-input}. Those of the proton and deuteron atomic masses are: $\ArMCODATA{p} = 1.007\, 276\, 466\, 621$ and $\ArMCODATA{d} = 2.013\, 553\, 212\, 745$. The mass ratios $\lambda_{\rm pd}$ and $\lambda_{\rm d}$ are calculated using $\lambda_{\rm pd} = \ArM{p} \ArM{d} / ( \ArM{e}[\ArM{p} + \ArM{d}])$ and $\lambda_{\rm d} = \ArM{d}/\ArM{e}$ (and similar for the reference values $\lambda_{\rm pd0}$ and $\lambda_{\rm d0}$).

As explained above, we furthermore have observational equations of the form
\begin{equation}
\delta f(v,L,v',L') \doteq \delta f^{\rm th}(v,L,v',L'),
\end{equation}
with input value $\delta f(v,L,v',L') = 0$, to deal with theoretical uncertainties.

Finally, we require observational equations for $R_{\infty}$, $r_{\rm p}$, $r_{\rm d}$ as well as relative atomic masses and mass ratios. They are written as
\begin{subequations}
\begin{gather}
R_{\infty}^{\rm CODATA2018} \doteq R_{\infty} \,, \\
r_{\rm p}^{\rm CODATA2018} \doteq r_{\rm p} \,,
\end{gather}
\end{subequations}
etc., where the input data $R_{\infty}^{\rm CODATA2018}$, $r_{\rm p}^{\rm CODATA2018}$, $\ldots$ are found in Table~\ref{table-fc-input} under labels C1-C7 (note that only C5-C7 are used in Adjustment 2). The correlation coefficients among these input data are given in Table~\ref{table-correl}.

The coefficients $\beta$, $\eta$ in Eq.~(\ref{obs-eq-adj2}) depend on $(v,L,v',L')$, but for brevity this dependence is not indicated. $\beta_{R_{\infty}}$ is the linear sensitivity coefficient to the Rydberg constant, and is simply given by
\begin{equation}
\beta_{R_{\infty}} (v,L,v',L') = \frac{f_{\rm SA}^{\rm th} [(v,L)\to (v',L')]}{c R_{\infty}}.
\end{equation}
In the dependence on the nuclear charge radii, following CODATA practice we only take into account the leading-order finite-size correction~(see Eq.~(6) of~\cite{Korobov2006}). Higher-order finite-size corrections for the deuteron, which are included in the theoretical predictions~\cite{Korobov2021}, are taken as fixed, and the corresponding uncertainties are taken into account in $\delta f$. This yields a quadratic dependence on $r_{\rm p}$ and $r_{\rm d}$, with sensitivity coefficients
\begin{subequations}
\begin{align}
\eta_{r_i} (v,L,v',L') &= \eta_{r_i} (v',L') - \eta_{r_i} (v,L), \\
\eta_{r_i} (v,L) &= (2cR_{\infty}) \frac{2 \pi a_0}{3} \left\langle \delta(\mathbf{x}_i) \right\rangle \, 
\end{align}
\end{subequations}
with $i = \text{p,d}$. $a_0$ is Bohr's radius, and $\mathbf{x}_{\rm p}$ ($\mathbf{x}_{\rm d}$) is the position vector of the electron with respect to the proton (deuteron). Brackets denote expectation values over the nonrelativistic wavefunction of the $(v,L)$ rovibrational state, which are calculated numerically with high precision using variational wavefunctions~\cite{Aznabayev2019}.

Finally, in order to evaluate the linear sensitivity coefficients with respect to the mass ratios, $\beta_{\lambda_{\rm pd}}$ and $\beta_{\lambda_{\rm d}}$, it suffices to consider the nonrelativistic energy levels, which yields a relative accuracy on the order of $\alpha^2$ for these coefficients. Following the approach of~\cite{Schiller2005}, they can be expressed as
\begin{subequations}
	\begin{align}
	\beta_{\lambda_i} (v,L,v',L') &= \beta_{\lambda_i} (v',L') - \beta_{\lambda_i} (v,L), \\
	\beta_{\lambda_{\rm pd}} (v,L) &= (2c R_{\infty}) \frac{a_0^2}{2\hbar\lambda_{\rm pd}^2} \left\langle \nabla_{\mathbf{X}}^2 \right\rangle, \\
	\beta_{\lambda_{\rm d}} (v,L) &= (2c R_{\infty}) \frac{a_0^2}{\hbar\lambda_{\rm d}^2} \left( \frac{1}{2} \left\langle \nabla_{\mathbf{x}_{\rm d}}^2 \right\rangle  \!+\!  \left\langle \nabla_{\mathbf{X}} \cdot \nabla_{\mathbf{x}_{\rm d}} \right\rangle \right),
	\end{align}
\end{subequations}
where $\mathbf{X}$ is the position of the proton with respect to the deuteron. The expectation values are calculated numerically with high precision using variational wavefunctions~\cite{Schiller2005}.

For Adjustment 3, the mass dependence is parametrised in terms of the relative atomic masses, \Ar{e}, \Ar{p}, and \Ar{d}. The observational equation~(\ref{obs-eq-adj2}) becomes
\begin{equation} \label{obs-eq-adj3}
\begin{split}
&\,f_{\rm SA} [(v,L)\to (v',L')] \\
&\doteq f_{\rm SA}^{\rm th} [(v,L)\to (v',L')] + \delta f(v,L,v',L') \\
&\quad + \beta_{R_{\infty}} c(R_{\infty} \!-\! R_{\infty 0}) + \eta_{r_{\rm p}} (r_{\rm p}^2 \!-\! r_{{\rm p}0}^2) + \eta_{r_{\rm d}} (r_{\rm d}^2 \!-\! r_{{\rm d}0}^2) \\
&\quad + \beta_{\ArM{e}} (\ArM{e} \!-\! \ArM{e}_0) + \beta_{\ArM{p}} (\ArM{p} \!-\! \ArM{p}_0) + \beta_{\ArM{d}} (\ArM{d}) \!-\! \ArM{d}_0) \,.
\end{split}
\end{equation}
Here, the observational equations, values, and correlation coefficients for $\delta f$ are the same as for Adjustment 2. The sensitivity coefficients $\beta_{\ArM{\it i}}$ ($i={\rm e,p,d}$) with respect to the relative atomic masses are given by
\begin{equation} \label{eq:betaconversion}
\beta_{\ArM{\it i}} (v,L,v',L') = \frac{\partial\lambda_{\rm pd}}{\partial \ArM{\it i}} \beta_{\lambda_{\rm pd}} (v,L,v',L') + \frac{\partial\lambda_{\rm d}}{\partial \ArM{\it i}} \beta_{\lambda_{\rm d}} (v,L,v',L').
\end{equation}
The theoretical frequencies and all the sensitivity coefficients appearing in Eqs.~(\ref{obs-eq-adj2}) and (\ref{obs-eq-adj3}) are given in Table~\ref{table-sensit}. The dependence of the sensitivity coefficients on fundamental constants is weak, so that the values of Table~\ref{table-sensit} can be safely used during all iterations of the least-squares adjustment procedure. Note that the theoretical frequencies differ very slightly from those given in~\cite{Korobov2021} because they are calculated with 2018 CODATA values of $\ArM{e}$, $\ArM{p}$ and $\ArM{d}$, whereas the values of~\cite{Korobov2021} were obtained with 2018 CODATA values of \mpme{} and \mdme{}. When switching between these two parametrizations, round-off errors in the published recommended values cause tiny shifts in the nonrelativistic energy levels.

In total, Adjustment 2 contains $N = 9$ input data and $M = 7$ adjusted parameters: $\lambda_{\rm pd}$, $R_{\infty}$, $r_{\rm p}$, $r_{\rm d}$, and the additive corrections for the three rovibrational transition frequencies.

Adjustment 3 combines the input data of the first two adjustments, with $N = 13$ input data and $M = 9$ adjusted parameters.

\begin{table*}\scalebox{0.95}{
\begin{tabular}{@{\hspace{2mm}}c@{\hspace{3mm}}c@{\hspace{3mm}}c@{\hspace{3mm}}c@{\hspace{3mm}}c@{\hspace{2mm}}} 
\hline
Transition & $f_{\rm SA}^{\rm theor}$ & $\beta_{cR_{\infty}}$ & $\eta_{r_p}$    & $\eta_{r_d}$ \\
           & (kHz)                    &                       & (kHz.fm$^{-2}$) & (kHz.fm$^{-2}$) \\
\hline \hline
$(0,0) \to (0,1)$ &  1 314 925 752.929  & 3.9969[-04] & -9.0991[-01] & -9.0991[-01] \\
$(0,0) \to (1,1)$ & 58 605 052 163.88   & 1.7814[-02] & -2.4253[+01] & -2.4220[+01] \\
$(0,3) \to (9,3)$ & 415 264 925 502.7   & 1.2623[-01] & -1.5940[+02] & -1.5850[+02] \\
\hline
\end{tabular}}
\end{table*}
\vspace{-1cm}
\begin{table*}\scalebox{0.95}{ 
\begin{tabular}
{@{\hspace{2mm}}c@{\hspace{3mm}}c@{\hspace{3mm}}c@{\hspace{3mm}}c@{\hspace{3mm}}c@{\hspace{3mm}}c@{\hspace{2mm}}}
Transition & $\beta_{\lambda_{pd}}$ & $\beta_{\lambda_{d}}$ & $\beta_{\ArM{e}}$  & $\beta_{\ArM{p}}$  & $\beta_{\ArM{d}}$ \\
           & (kHz)                  & (kHz)                 & (kHz.u$^{-1}$) & (kHz.u$^{-1}$) &
(kHz.u$^{-1}$) \\
\hline \hline
$(0,0) \to (0,1)$ & -1.0601[+06] & -3.2126[+01] & 2.3653[+12] & -8.5857[+08] & -2.1492[+08] \\
$(0,0) \to (1,1)$ & -2.3201[+07] & -7.0874[+02] & 5.1767[+13] & -1.8790[+10] & -4.7036[+09] \\
$(0,3) \to (9,3)$ & -1.2580[+08] & -3.9569[+03] & 2.6664[+14] & -9.6785[+10] & -2.4227[+10] \\
\hline
\end{tabular}}
\caption{Reference theoretical values and sensitivity coefficients of \hdplus{} transition frequencies. The theoretical frequencies are calculated using 2018 CODATA values of $\ArM{e}$, $\ArM{p}$, $\ArM{d}$, $cR_{\infty}$, $r_p$, $r_d$, and $\alpha$. The sensitivity coefficient for the Rydberg constant $\beta_{cR_{\infty}}$ is dimensionless. u stands for the relative atomic mass unit. \label{table-sensit}}
\end{table*}

\begin{table*}\scalebox{0.9}{
	\begin{tabular}{@{\hspace{2mm}}c@{\hspace{2mm}}l@{\hspace{4mm}}l@{\hspace{4mm}}l@{\hspace{4mm}}c@{\hspace{2mm}}} 
		\hline
		Label & Input datum & Value & Rel. Uncertainty & Reference \\
		\hline
		A1 & $f_{\rm SA} [(0\!,\!0)\! \to\! (0\!,\!1)]$ & 1 314 925 752.978(48) kHz  & $3.7 \times 10^{-11}$ & this work \\
		A2 & $f_{\rm SA} [(0\!,\!0)\! \to\! (1\!,\!1)]$ & 58 605 052 164.14(56) kHz  & $9.6 \times 10^{-12}$ & this work \\
		A3 & $f_{\rm SA} [(0\!,\!3)\! \to\! (9\!,\!3)]$ & 415 264 925 501.3(1.6) kHz & $3.9 \times 10^{-12}$ & this work \\
		\hline
		B1 & $\delta f(0,0,0,1)$ & 0.000(19) kHz & & \cite{Korobov2021} \\
		B2 & $\delta f(0,0,1,1)$ & 0.00(49)  kHz & & \cite{Korobov2021} \\
		B3 & $\delta f(0,3,9,3)$ & 0.0(3.2)  kHz & & \cite{Korobov2021} \\
		\hline
		C1 & $\ArM{e}$ (CODATA 2018)      & 5.485 799 090 65(16)$\times 10^{-4}$ u & $2.9 \times 10^{-11}$ & \cite{Tiesinga2021} \\ 
		C2 & $\ArM{d}$ (AMDC 2020)        & 2.013 553 212 537(15) u                & $7.5 \times 10^{-12}$ & \cite{Wang:2021xhn} \\
		C3 & $\ArM{p}$ (AMDC 2020)        & 1.007 276 466 587(14) u                & $1.4 \times 10^{-11}$ & \cite{Wang:2021xhn} \\   
		C4 & \mdmp{} (Fink 2021)         & 1.999 007 501 272(9)                   & $4.5 \times 10^{-12}$ & \cite{Fink2021} \\
		C5 & $cR_{\infty}$ (CODATA 2018) & 3 289 841 960 250.8(6.4) kHz           & $1.9 \times 10^{-12}$ & \cite{Tiesinga2021} \\
		C6 & $r_{\rm p}$ (CODATA 2018)         & 0.8414(19) fm                          & $2.2 \times 10^{-3}$ & \cite{Tiesinga2021} \\
		C7 & $r_{\rm d}$ (CODATA 2018)         & 2.12799(74) fm                         & $3.5 \times 10^{-4}$ & \cite{Tiesinga2021} \\
		\hline
	\end{tabular}}
	\caption{Input data for the adjustments discussed in Section~\ref{sec-discussion}. Entries C1 and C5-C7 also serve as reference values in equations~(\ref{obs-eq-adj2}) and (\ref{obs-eq-adj3}). \label{table-fc-input}}
\end{table*}

\begin{table*}\scalebox{0.9}{
	\begin{tabular}{@{\hspace{2mm}}l@{\hspace{3mm}}l@{\hspace{3mm}}l@{\hspace{3mm}}l@{\hspace{2mm}}} 
		\hline
		\multicolumn{4}{c}{Correlation coefficients} \\
		\hline \hline
		$r({\rm A1},{\rm A2}) = 0.00036$ & $r({\rm A1},{\rm A3}) = 0.00644$ & $r({\rm A2},{\rm A3}) =   
         0.00741$ & $r({\rm B1},{\rm B2}) = 0.99570$ \\
        $r({\rm B1},{\rm B3}) = 0.95733$ & $r({\rm B2},{\rm B3}) = 0.97998$ & $r({\rm C1},{\rm C5}) = 0.00704$ & $r({\rm C1},{\rm C6}) = -0.00133$ \\
        $r({\rm C1},{\rm C7}) = 0.00317$ & $r({\rm C2},{\rm C3}) = 0.31986$ & $r({\rm C5},{\rm C6}) = 0.88592$ & $r({\rm C5},{\rm C7}) = 0.90366$ \\ 
        $r({\rm C6},{\rm C7}) =0.99165$ & & & \\
		\hline
	\end{tabular}}
	\caption{Values of all nonzero correlation coefficients in the input data of Table~\ref{table-fc-input}. \label{table-correl}}
\end{table*}

\vspace{5mm}

\subsection{Results and discussion}

\begin{table*}
	\begin{tabular}{@{\hspace{2mm}}c@{\hspace{2mm}}c@{\hspace{2mm}}c@{\hspace{2mm}}c@{\hspace{2mm}}}
		\hline
		Adjustment & Input data & Adjusted value & $u_{\rm r}$ ($10^{-11}$) \\
		\hline
		1 & C1-C4           & $\lambda_{\rm pd} = 1 \, 223.899 \, 228 \, 646(37)$ & 3.0 \\
		2 & A,B,C5-C7       & $\lambda_{\rm pd} = 1 \, 223.899 \, 228 \, 719(26)$ & 2.1 \\
        1 & C1-C4           & $\lambda_{\rm p} = 1 \, 836.152 \, 673 \, 353(56)$    & 3.0 \\
		3 & All             & $\lambda_{\rm p} = 1 \, 836.152 \, 673 \, 423(33)$    & 1.8 \\
        3 & All             & $\ArM{e} = 5.485 \, 799 \, 090 \, 46(10) \times 10^{-4}$~u & 1.8 \\
		\hline
	\end{tabular}
	\caption{ Adjustment results. $u_{\rm r}$ stands for relative uncertainty.\label{table-adj2-results}}
\end{table*}
The determinations of $\lambda_{\rm pd}$ and of the proton-electron mass ratio, $\lambda_{\rm p}$, from the adjustments described above are shown in Table~\ref{table-adj2-results}.

Adjustment 1 illustrates the potential precision permitted by Penning trap measurements alone. The relative uncertainty of $\lambda_{\rm pd}$ is found to be $3.0 \times 10^{-11}$, mainly limited by the electron's relative atomic mass that has a relative uncertainty of $2.9 \times 10^{-11}$ (see Table~\ref{table-fc-input}). The contribution from the proton and deuteron relative atomic masses to the uncertainty of $\lambda_{\rm pd}$ is $7.6 \times 10^{-12}$.

Adjustment 2 provides an alternative determination of $\lambda_{\rm pd}$ from \hdplus{} spectroscopic data. Its uncertainty is slightly smaller ($2.1 \times 10^{-11}$), and the obtained values are in reasonable agreement, with a mild (1.6$\sigma$) tension. Since the precision of the first adjustment is limited by the electron's atomic mass, the smaller uncertainty from \hdplus{} data indicates that the impact of \hdplus{} on the global CODATA adjustment would primarily be to improve the precision of $\ArM{e}$.

This is verified in the last adjustment (Adjustment 3), which illustrates the potential improvement that can be obtained by combining Penning trap with \hdplus{} spectroscopy data. Whereas the proton and deuteron atomic masses are very precisely determined by mass spectrometry, \hdplus{} spectroscopy, being sensitive to the nucleus-to-electron mass ratios, provides a link to the electron mass, allowing to improve the precision of $\ArM{e}$ and \mpme{} to $1.8 \times 10^{-11}$, which for the latter would represent an improvement by more than a factor of three over its 2018 CODATA uncertainty. It is worth noting that the adjustment reveals no significant tension in the data, with all normalised residuals being smaller than 1.2 in absolute value. The found values of $\ArM{e}$ and \mpme{} are furthermore in good agreement with the 2018 CODATA values.

We carried out Adjustment 3 taking into account the weak correlations between the adjusted experimental spin-averaged frequencies, discussed in Section~\ref{sec-globaladjC}. To assess the influence of these correlations, we repeated Adjustment 3 assuming zero correlation between the spin-averaged frequencies. The value of \mpme{} obtained from this adjustment is identical to the one shown in the last lines of Table~\ref{table-adj2-results}, indicating a negligible impact of these correlations on the determination of fundamental constants from recent \hdplus{} theory and measurements.

\section{Conclusion} \label{sec-concl}
In conclusion, we have carried out a comprehensive analysis of the available high-precision spectroscopic data of rotational and vibrational transitions in \hdplus{}, with the aim of providing a set of spin-averaged rotational-vibrational transition frequencies that can serve as input data for future CODATA adjustments of the fundamental physical constants. A crucial part of the analysis involves the use of theoretically computed hyperfine structure to remove hyperfine shifts from the measured transition frequencies, taking into account the correlations between theroretical spin coefficients. The experimental input data are complemented by the most recent theoretical predictions of the spin-averaged transition frequencies, and by linearised expressions that parametrise these theoretical predictions in terms of relevant fundamental constants. Uncertainties and correlation coefficients pertaining to the input data were also provided, which completes the set of observational equations and covariances needed for future CODATA adjustments. To illustrate the potential of the \hdplus{} data, we have carried out adjustments of the electron, proton and deuteron relative atomic mass using state-of-the-art data from Penning-trap measurements as well as the \hdplus{} data presented here. We found that the \hdplus{} data have a particularly large impact on the precision of the electron relative atomic mass which, together with the precise values of the proton and deuteron relative atomic masses from Penning-trap measurements, may improve the precision of \mpme{} by more than a factor of three compared to the current 2018 CODATA value.

\section*{Acknowledgement(s)}
The authors warmly thank Wim Ubachs for his invaluable contributions and continued support to the Amsterdam-Paris HD$^+$ research collaboration. Eite Tiesinga and Fabian Hei{\ss}e are acknowledged for helpful comments to the manuscript.

\section*{Disclosure statement}
The authors report there are no competing interests to declare.

%

%

%

%


\bibliographystyle{tfo}
\bibliography{spinaveraged_MolPhys_v0}

\providecommand{\noopsort}[1]{}\providecommand{\singleletter}[1]{#1}%
\begin{thebibliography}{39}
\providecommand{\url}[1]{\texttt{#1}}
\providecommand{\urlprefix}{URL }

\bibitem{Tiesinga2021}
E. Tiesinga, P.J. Mohr, D.B. Newell and B.N. Taylor,  Rev. Mod. Phys.
  \textbf{93}, 025010 (2021).

\bibitem{Sturm2014}
S. Sturm, F. K\"ohler, J. Zatorski, A. Wagner, Z. Harman, G. Werth, W. Quint,
  C.H. Keitel and K. Blaum,  Nature  \textbf{506}, 467--470 (2014).

\bibitem{Heisse2017}
F. Hei{\ss}e, F. K{\"o}hler-Langes, S. Rau, J. Hou, S. Junck, A. Kracke, A.
  Mooser, W. Quint, S. Ulmer, G. Werth, K. Blaum and S. Sturm,  {Phys. Rev.
  Lett.}  \textbf{119} (3), 033001 (2017).

\bibitem{Heisse2019}
F. Hei\ss{}e, S. Rau, F. K\"ohler-Langes, W. Quint, G. Werth, S. Sturm and K.
  Blaum,  Phys. Rev. A  \textbf{100}, 022518 (2019).

\bibitem{Rau2020}
S. Rau, F. Hei{\ss}e, S. K{\"o}hler-Langes, F.~Sasidharan, R. Haas, D. Renisch,
  C.E. D{\"u}llmann, W. Quint, S. Sturm and K. Blaum,  {Nature}  \textbf{585},
  43--47 (2020).

\bibitem{Fink2020}
D.J. Fink and E.G. Myers,  Phys. Rev. Lett.  \textbf{124}, 013001 (2020).

\bibitem{Fink2021}
D.J. Fink and E.G. Myers,  Phys. Rev. Lett.  \textbf{127}, 243001 (2021).

\bibitem{Wing1976}
W.H. Wing, G.A. Ruff, W.E. Lamb and J.J. Spezeski,  Phys. Rev. Lett.
  \textbf{36}, 1488--1491 (1976).

\bibitem{Hori2006}
M. Hori, A. Dax, J. Eades, K. Gomikawa, R.S. Hayano, N. Ono, W. Pirkl, E.
  Widmann, H.A. Torii, B. Juh\'asz, D. Barna and D. Horv\'ath,  Phys. Rev.
  Lett.  \textbf{96}, 243401 (2006).

\bibitem{Hori2011}
M. Hori, A. S\'ot\'er, D. Barna, A. Dax, R. Hayano, S. Friedreich, B. Juh\'asz,
  T. Pask, E. Widmann, D. Horv\'ath, L. Venturelli and N. Zurlo,  Nature
  \textbf{475}, 484--488 (2011).

\bibitem{Hori2016}
M. Hori, H. Aghai-Khozani, A. S\'ot\'er, D. Barna, A. Dax, R. Hayano, T.
  Kobayashi, Y. Murakami, K. Todoroki, H. Yamada, D. Horv\'ath and L.
  Venturelli,  Science  \textbf{354}, 610--614 (2016).

\bibitem{Mohr2008}
P.J. Mohr, B.N. Taylor and D.B. Newell,  Rev. Mod. Phys.  \textbf{80}, 633--730
  (2008).

\bibitem{Mohr2012}
P.J. Mohr, B.N. Taylor and D.B. Newell,  Rev. Mod. Phys.  \textbf{84},
  1527--1605 (2012).

\bibitem{Alighanbari2020}
S. Alighanbari, G.S. Giri, F.L. Constantin, V.I. Korobov and S. Schiller,
  {Nature}  \textbf{581}, 152--158 (2020).

\bibitem{Patra2020}
S. Patra, M. Germann, J.{\relax -Ph}. Karr, M. Haidar, L. Hilico, V.I. Korobov,
  F.M.J. Cozijn, K.S.E. Eikema, W. Ubachs and J.C.J. Koelemeij,  Science
  \textbf{369} (6508), 1238--1241 (2020).

\bibitem{Kortunov2021}
I.V. Kortunov, S. Alighanbari, M.G. Hansen, G.S. Giri, V.I. Korobov and S.
  Schiller,  Nat. Phys.  \textbf{17}, 569--573 (2021).

\bibitem{Korobov2017}
V.I. Korobov, L. Hilico and J.{\relax -Ph}. Karr,  {Phys. Rev. Lett.}
  \textbf{118}, 233001 (2017).

\bibitem{Aznabayev2019}
D.T. Aznabayev, A.K. Bekbaev and V.I. Korobov,  Phys. Rev. A  \textbf{99},
  012501 (2019).

\bibitem{Korobov2021}
V.I. Korobov and J.{\relax -Ph}. Karr,  Phys. Rev. A  \textbf{104}, 032806
  (2021).

\bibitem{Karr2016}
J.{\relax -Ph}. Karr, L. Hilico, J.C.J. Koelemeij and V.I. Korobov,  Phys. Rev.
  A  \textbf{94}, 050501(R) (2016).

\bibitem{Karshenboim2017}
S.G. {Karshenboim} and V.G. {Ivanov},  Applied Physics B: Lasers and Optics
  \textbf{123} (1), 18 (2017).

\bibitem{Antognini2022}
A. Antognini, F. Hagelstein and V. Pascalutsa,  Annual Review of Nuclear and
  Particle Science  \textbf{72} (1), 389--418 (2022).

\bibitem{Antognini2022arXiv}
A. Antognini, S. Bacca, A. Fleischmann, L. Gastaldo, F. Hagelstein, P.
  Indelicato, A. Knecht, V. Lensky, B. Ohayon, V. Pascalutsa, N. Paul, R. Pohl
  and F. Wauters, {Muonic-Atom Spectroscopy and Impact on Nuclear Structure and
  Precision QED Theory. arXiv:2210.16929 [nucl-th]} arXiv 2022.
  $<${https://arxiv.org/abs/2210.16929}$>$.

\bibitem{Koelemeij2022}
J.C.J. Koelemeij,  Molecular Physics  \textbf{120} (19-20), e2058637 (2022).

\bibitem{Korobov22}
V.I. Korobov,  Phys. Part. Nucl.  \textbf{53}, 787 (2022).

\bibitem{Bakalov06}
D. Bakalov, V.I. Korobov and S. Schiller,  Phys. Rev. Lett.  \textbf{97},
  243001 (2006).

\bibitem{Karr2020}
J.{\relax -Ph}. Karr, M. Haidar, L. Hilico, Z.X. Zhong and V.I. Korobov,
  {Phys. Rev. A}  \textbf{102}, 052827 (2020).

\bibitem{Layzer64}
A. Layzer,  Nuovo Cim.  \textbf{33}, 1538 (1964).

\bibitem{Zwanziger64}
D.E. Zwanziger,  Nuovo Cim.  \textbf{34}, 77 (1964).

\bibitem{Bodwin88}
G.T. Bodwin and D.R. Yennie,  Phy. Rev. D  \textbf{37}, 498 (1988).

\bibitem{Korobov20}
V.I. Korobov, J.P. Karr, M. Haidar and Z.X. Zhong,  Phys. Rev. A  \textbf{102},
  022804 (2020).

\bibitem{Haidar22a}
M. Haidar, V.I. Korobov, L. Hilico and J.P. Karr,  Phys. Rev. A  \textbf{106},
  022816 (2022).

\bibitem{Haidar22b}
M. Haidar, V.I. Korobov, L. Hilico and J.P. Karr,  Phys. Rev. A  \textbf{106},
  042815 (2022).

\bibitem{Puchalski20}
M. Puchalski, J. Komasa and K. Pachucki,  Phys. Rev. Lett.  \textbf{125}
  (2020).

\bibitem{Mohr2000}
P.J. Mohr and B.N. Taylor,  Rev. Mod. Phys.  \textbf{72}, 351--495 (2000).

\bibitem{Wang:2021xhn}
M. Wang, W.J. Huang, F.G. Kondev, G. Audi and S. Naimi,  Chin. Phys. C
  \textbf{45} (3), 030003 (2021).

\bibitem{Huang:2021nwk}
W.J. Huang, M. Wang, F.G. Kondev, G. Audi and S. Naimi,  Chin. Phys. C
  \textbf{45} (3), 030002 (2021).

\bibitem{Korobov2006}
V.I. Korobov,  Phys. Rev. A  \textbf{74}, 052506 (2006).

\bibitem{Schiller2005}
S. Schiller and V. Korobov,  Phys. Rev. A  \textbf{71}, 032505 (2005).

\end{thebibliography}

%
%

\appendix
\renewcommand{\thefigure}{A\arabic{figure}}
\setcounter{figure}{0}
\section{Monte-Carlo simulation} \label{app-monte-carlo}
\begin{figure*}[b]
	\centering \includegraphics[width=1\columnwidth]{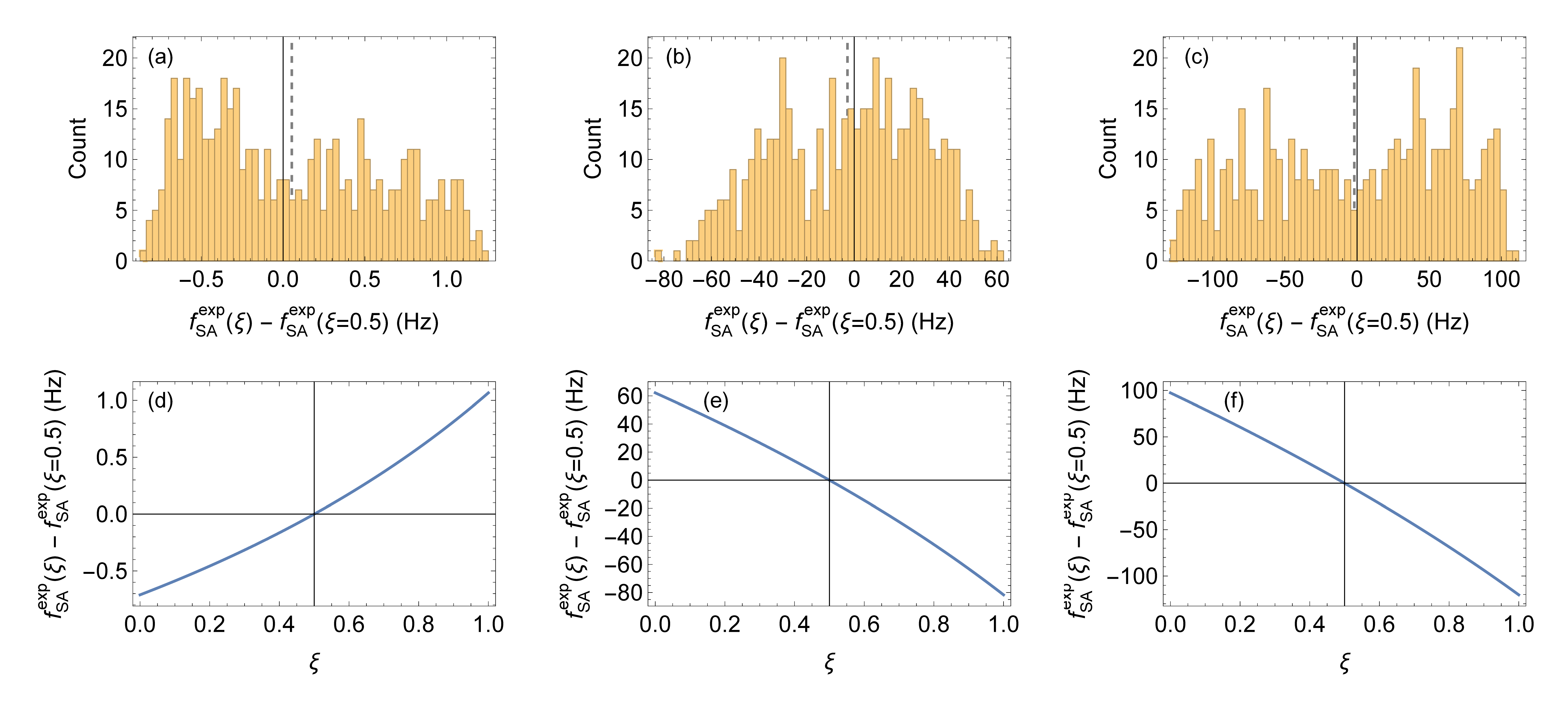}
	\caption{Distribution of correlation-induced shifts of spin-averaged frequencies obtained from a Monte-Carlo simulation consisting of 502 runs. Here the components of the vector $(r_1,r_2,r_6,r_8,r_9)$ are varied together as $(\xi,\xi,\xi,\xi,\xi)$; for example, $\xi=0.5$ stands for the vector $(0.5,0.5,0.5,0.5,0.5)$ (a) Distribution for the \vRot{} transition. (b) Distribution for the \vVib{} transition. (c) Distribution for the \vTP{} transition. Dashed vertical lines indicate the mean value of the frequency distribution. (d) Frequency shift of the \vRot{} transition versus $\xi$, with all components of the vector $(r_1,r_2,r_6,r_8,r_9)$ being varied together as $(\xi,\xi,\xi,\xi,\xi)$. (e) Same as (d), but for the \vVib{} transition. (f) Same as (d), but for the \vTP{} transition.} \label{fig-histo}
\end{figure*}

To assess the uncertainty associated with the unknown correlation coefficients $r[E_k(v,L),E_k(v',L')]\equiv r_{k}$ (with $k\neq 4,5$) we carry out a Monte-Carlo simulation analogous to \cite{Koelemeij2022}. As explained in Section~\ref{sec-globaladjA}, we set $r_2=r_3$ and $r_6=r_7$. We subsequently choose random values from the interval $[0,1)$ for each of the coefficients of the vector $(r_1,r_2,r_6,r_8,r_9)$, and carry out the adjustment of the spin-averaged frequencies and $\delta E_k (v\!,\!L)$. This is repeated 500 times, and in addition for the vectors $(0,0,0,0,0)$ and $(1-\epsilon,1-\epsilon,1-\epsilon,1-\epsilon,1-\epsilon)$, with $\epsilon=10^{-6}$ to avoid singular covariance matrices. Adjustments are carried out numerically using 100 digits of precision. Histograms of the Monte-Carlo distributions of adjusted transition frequencies are shown in Figure~\ref{fig-histo}. The full widths of the frequency distributions are 2.09~Hz, 144~Hz, and 237~Hz for the \vRot{}, \vVib{} and \vTP{} transitions, respectively; the deviation of the mean spin-averaged frequency from the frequency obtained for the $(0.5,0.5,0.5,0.5,0.5)$ case is 0.05~Hz, $-3$~Hz, and $-2$~Hz (i.e. less than $3\%$ of the full width). Figure~\ref{fig-histo} furthermore shows the dependence of the correlation-induced shift if all correlation coefficients are varied as $(\xi,\xi,\xi,\xi,\xi)$, with $\xi$ between 0 and 1. Comparing the shifts of the three transitions shows that the shift of the rotational transition is anticorrelated with the shift of the $v=0 \to 1$ and $v=0 \to 9$ transitions, whereas the shifts for the latter two are correlated.

\end{document}